\makeatletter

\newcommand{\Rmnum}[1]{\expandafter\@slowromancap\romannumeral #1@}
\makeatother

\documentclass{ccjnl}
\usepackage{url}

\title{Space-Based Computing Networks: Trends, Architecture, Challenges, and Key Technologies.}
\author{Linling Kuang\inst{$\dagger$,2,3}, Jiachen Sun\inst{1}, Jin Zhang\inst{1}, Huanxi Cui\inst{2,3}, Kai Liu\inst{$\dagger$,2,3}}

\receiveddate{8th. March, 2025}
\reviseddate{23rd. July, 2025}
\Editor{XXX XXX}

\address[1]{Department of Electronic Engineering, Tsinghua University, Beijing 100084, China}
\address[2]{State Key Laboratory of Space Network and Communications, Tsinghua University, Beijing 100084, China}
\address[3]{Beijing National Research Center for Information Science and Technology, Tsinghua University, Beijing 100084, China}
\address[$\dagger$]{Corresponding author.}

\begin{document}
\maketitle

\begin{abstract}
As one of the most promising hotspots in the 6G era, space remote sensing information networks play a key and irreplaceable role in areas such as emergency response and scientific research, and are expected to foster remote sensing data processing into the next generation of killer applications. However, due to the inability to deploy ground communication stations at scale and the limited satellite-to-ground link rate, the traditional model for transmitting space data back to ground stations faces significant challenges in terms of timeliness. To address this problem, we focus on the emerging paradigm of on-orbit space data processing, which reduces the volume of transmitted data by several orders of magnitude to enable faster task response, taking the first step toward building a space-based computing network. Specifically, we propose a hierarchical space-based computing network architecture, comprising the space-based cloud constellation system, the remote sensing constellation system, the network operation control center, the orchestration data center, and the user access portal. Each component is described in detail from a system design perspective to clarify its specific role and functionality. Next, we analyze three scientific challenges: the heterogeneous resource virtualization and state information synchronization, the matching of multi-priority tasks with multidimensional resources, and the fault detection and localization under extreme conditions. Finally, we discuss key technologies to address the aforementioned challenges and highlight promising research priorities for the future.
\end{abstract}

\section{Introduction}

{With the advent of the 6G era, networks such as mobile communications, industrial control, and the Internet of Things are rapidly evolving, propelling industries toward greater intelligence and digitalization. In this landscape, space remote sensing information networks stand out for their wide coverage and efficient real-time data acquisition, enabling fast service for both production and daily life, and are expected to drive the next generation of groundbreaking applications \cite{yao2018space}\cite{1}\cite{navalgund2007remote}.}


{However, unlike terrestrial networks, the limited communication transmission capacity and the dynamically unstable connections inherent in space networks often lead to waiting times ranging from tens of minutes to several hours in the traditional model of transmitting space data back to the ground for processing, which significantly limits the rapid response capabilities
 \cite{sun2024distributionally}.} On the one hand, due to constraints such as geographic conditions and the high costs of manpower and resources, ground stations cannot be deployed in large numbers, leading to an inability to transmit space data from any location at any time \cite{baeza2023gateway}. On the other hand, most current remote sensing satellites are positioned in Low Earth Orbit (at altitudes ranging from several hundred to about a thousand kilometers) and move rapidly. The time window for data transmission to a ground station, when the satellite passes overhead, is typically only around ten minutes. As a result, transmitting tens of gigabytes of data back to the ground within this limited time frame is infeasible, given the constrained feeder link bandwidth in the Gbps range \cite{CHEN2023337}. Both of these challenges render the current model of transmitting space data to ground stations for processing inadequate to meet the rapid response demands of users.

Although both academia and industry are actively advancing research on high-speed satellite-to-ground transmission technologies, with the goal of achieving a significant improvement in transmission rates through resource optimization and multiplexing techniques \cite{li2024resource} \cite{chen2022trends}, the inherent limitations of the communication technologies themselves pose a substantial barrier that researchers are unlikely to overcome in the short term. Currently, satellite-to-ground data transmission primarily relies on two methods: microwave communication and free space optical (FSO) communication \cite{israel2018next}. For the former, microwave communication links are limited by their lower frequency bands, which restrict their maximum communication rate to the Gbps level, resulting in relatively low transmission speeds. As for the latter, although laser communication links offer transmission rates that are more than two orders of magnitude higher than those of microwave communication links, ranging from tens to hundreds of Gbps, they are highly susceptible to environmental factors such as clouds, fog, and turbulence, which hinder stable and continuous transmission, resulting in low reliability \cite{ippolito2017satellite}.

Therefore, to overcome the challenge of slow event response times, we shift away from traditional space data transmission models that rely on ground stations, adopting the emerging paradigm of on-orbit data processing, which reduces the volume of transmitted data by orders of magnitude. {Both industry and academia are actively pursuing research efforts to break through the key technologies required for on-orbit processing in space-based networks. Prominent companies such as Amazon \cite{aws}, Orbits Edge \cite{orbitsedge}, along with research institutions like the Chinese Academy of Sciences \cite{rs13020271} and Tsinghua University, have made significant progress in conducting technical validation. Amazon and Orbital Edge focus on deploying machine learning cloud services onboard satellites, successfully enabling on-orbit data processing that reduces image data volume by 42\% and significantly accelerates processing speed. The Aerospace Information Research Institute of the Chinese Academy of Sciences has extensive experience and a strong foundation in remote sensing data acquisition, compression, and processing, supported by the development of dedicated payloads. Experimental evaluations based on real-world datasets demonstrate that the system exhibits strong performance. Tsinghua University has been actively involved in the development of Medium Earth Orbit (MEO) satellite networks, which leverage their broad coverage capabilities to provide stable artificial intelligence services for multimodal data fusion and processing, aiming to address the urgent challenge of low-accuracy on-orbit data detection. Building on the successful launch of China’s first MEO broadband communication satellite, TSN (Tsinghua Satellite Network)-01 \cite{xhs}, the research team at Tsinghua University is advancing experiments in on-orbit processing for space-based data fusion, marking a significant step toward the establishment of a space-based computing network.} Space-based computing network is becoming an essential path for the future development of space information networks, and is expected to become a frontier in international research. 

\begin{figure*}
	\centering
	\includegraphics[width=1\textwidth, height=0.46\textheight]{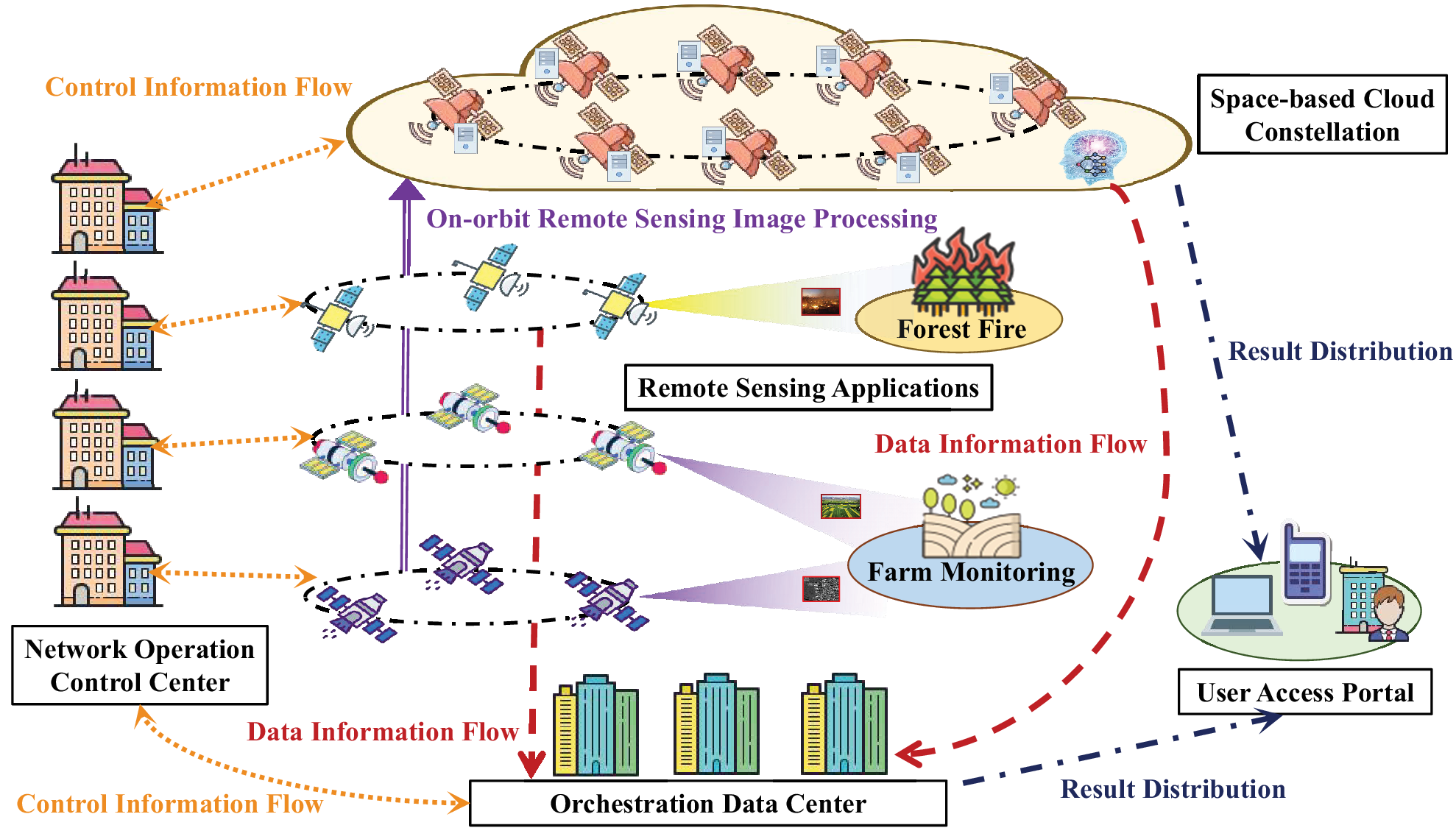}
	\caption{Hierarchical Space-Based Computing Network Architecture.}
\end{figure*}

Nevertheless, current research on space-based computing mainly focuses on individual technologies, such as optimizing AI algorithms to improve solution speed and recognition accuracy \cite{9676638} \cite{9437634}, or designing payload software structures to shield underlying interface differences and mitigate heterogeneity \cite{7302778} \cite{Ding_2022}. Building hierarchical space-based computing networks and analyzing related challenges and technologies from a global system perspective remains an untapped opportunity. Accordingly, our team will present our vision for space-based computing networks, with a focus on four key aspects: network architecture, challenge analysis, key technologies, and innovative practices.

The remainder of this paper is organized as follows. In Section \Rmnum{2}, we introduce the proposed hierarchical space-based computing network architecture, which involves the space-based cloud constellation system, the remote sensing constellation system, the network operation control center, the orchestration data center and the user access portal. Section \Rmnum{3} analyzes the challenges related to space-based computing networks in detail. Section \Rmnum{4} explores key technologies urgently requiring research in this field. Section \Rmnum{5} concludes this paper.


\section{Hierarchical Space-Based Computing Network Architecture}

\begin{figure*}
	\centering
	\includegraphics[scale=0.5]{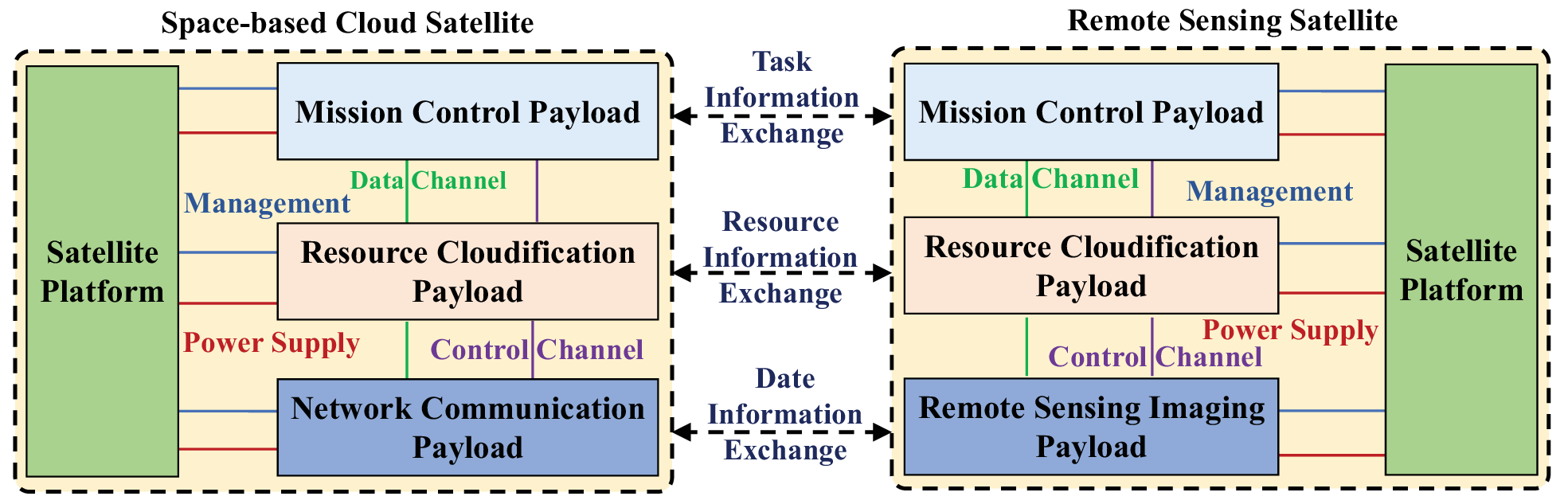}
	\caption{Satellite Configuration.}
\end{figure*}


As a highly successful satellite service provider in the world, SpaceX's Starlink project has gained global trust and recognition from numerous users and organizations, supported by its 7,316 satellites in orbit (as of December 5, 2024) \cite{numberofsatellites} and over 150 ground stations \cite{numberofgss}. Starlink employs a traditional architecture that relies on ground-based internet as its backbone, using satellites as relay nodes to forward space data to ground-based cloud centers for processing \cite{starlinksys1} \cite{starlinksys2}. However, as previously analyzed, this architecture is not the preferred solution due to its high manpower and resource costs, as well as its limitations in supporting the real-time transmission of large volumes of space data, such as those generated by remote sensing applications.

{Therefore, we propose a hierarchical space-based computing network architecture, in which data and control flows are decoupled to improve system efficiency, as illustrated in Figure 1. Unlike the Starlink model, which transmits space data to the ground for processing, we develop a space-based cloud constellation with powerful computational capabilities to support on-orbit processing of spatial information, thereby laying a solid foundation for the realization of an infrastructure-as-an-intelligent-service network. The network architecture comprises five components: the space-based cloud constellation system, the remote sensing constellation system, the network operation control center, the orchestration data center and the user access portal, each described in detail below.}

{\textbf{Space-based Cloud Constellation System (SCCS)}: The SCCS is not a single, fixed satellite constellation, but a collaborative architecture composed of Geosynchronous Earth Orbit (GEO), Medium Earth Orbit (MEO), and Low Earth Orbit (LEO) satellites equipped with computing payloads. GEO and MEO satellites, owing to their larger size and higher altitudes, can accommodate multiple computing payloads and provide wide-area coverage, making them well suited for delivering stable computational services for asynchronous, temporally staggered, and multimodal data fusion tasks. As illustrated in Figure 1, remote sensing satellites capturing images of the same farmland region at different times can rapidly upload data to the space-based cloud constellation for on-orbit processing. In the space environment, LEO satellites are the most numerous and typically function as edge computing nodes. Although their computational capabilities are generally inferior to those of GEO and MEO satellites, their performance can be significantly enhanced through high-speed inter-satellite laser links, which in some cases enable superior overall efficiency.

{\textbf{Remote Sensing Constellation System (RSCS)}: Remote sensing satellites, which are diverse in type and widely distributed, are the primary source of vast amounts of data for space information networks and play a crucial role in scientific exploration and emergency disaster relief. The most advanced remote sensing satellites can be equipped with onboard computing payloads. However, since space-based power supply relies solely on solar energy and remote sensing imaging is highly energy-intensive, performing large-scale image recognition remains challenging. Consequently, onboard processing in RSCS is typically limited to preprocessing tasks such as radiometric correction, geometric correction, and noise reduction. As shown in Figure~1, when a remote sensing satellite patrols forested areas for fire detection, it can transmit the preprocessed images to the space-based cloud constellation for image recognition. For non-urgent tasks, remote sensing satellites can also transmit raw data in stages to the orchestration data center.

{\textbf{Network Operation Control Center (NOCC)}: The NOCC is typically associated with a specific constellation, responsible not only for the daily monitoring of the constellation's operational status but also for issuing detailed instructions, such as how communication links should be established and how computational resources should be allocated. It is a core component in managing the control information flows.}

{\textbf{Orchestration Data Center (ODC)}: The ODC serves as both a data reception hub and a task planner at the constellation level. On one hand, it receives data from the remote sensing constellation and the space-based cloud constellation, verifies the processing results, and distributes the outputs to end users. On the other hand, it performs coarse-grained task planning based on the current task volume from users and constellation load status provided by the NOCC. Specifically, the ODC determines how tasks should be assigned across constellations and transmits these high-level directives to the NOCC for execution. In contrast, fine-grained resource allocation—such as assigning specific satellites and configuring bandwidth or compute units—is handled internally by the NOCC.

{\textbf{User Access Portal (UAP)}: Through the UAP, users can submit task requests at any time and from anywhere, and receive feedback from the SCCS or the ODC.}

\begin{figure*}
	\centering
	\includegraphics[scale=0.5]{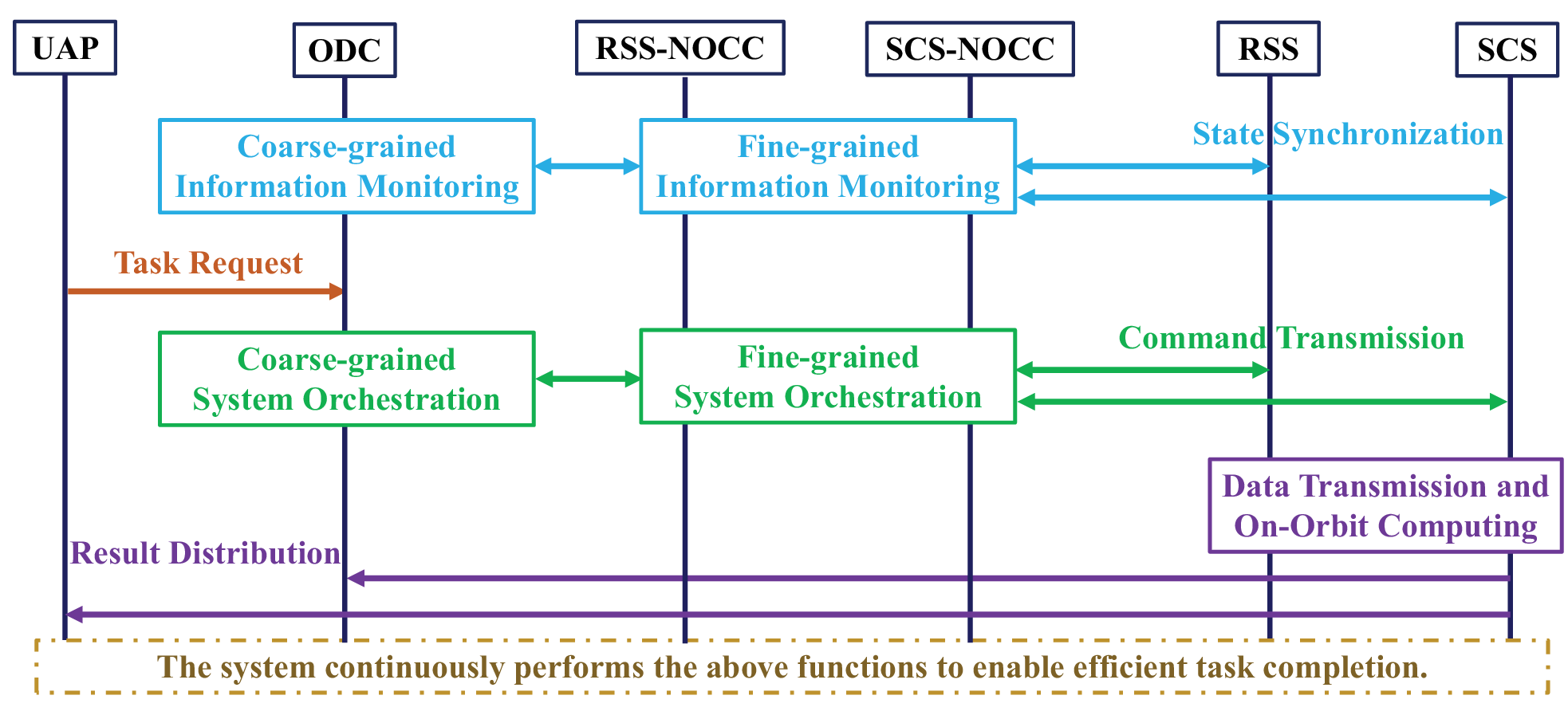}
	\caption{Workflow of Space-Based Computing Network.}
\end{figure*}

Furthermore, Figure 2 illustrates our satellite configuration design using an example of interaction between a typical remote sensing satellite (RSS) and a space-based cloud satellite (SCS). Whether RSS or SCS, their configurations primarily include task control payloads, resource cloudification payloads, and specialized functional payloads, such as network communication and remote sensing imaging. The task control payload facilitates task information exchange to support collaborative inter-satellite operations. The resource cloudification payload addresses challenges posed by heterogeneous resources—such as computing, storage, communication links, and data—across different satellites, which hinder interoperability. By abstracting and cloudifying these resources to mask interface differences, it efficiently supports complex tasks, such as multi-source data fusion processing. The specialized functional payload is determined by the satellite type and is primarily used for processing foundational data and facilitating interactions.

{Figure~3 presents the general workflow of the proposed space-based computing network architecture. RSS and SCS must maintain continuous, real-time status synchronization with their respective NOCC. Both the ODC and the NOCC utilize system information to support decision-making, but at different levels of granularity. The NOCC has comprehensive visibility into satellite operations, resources, and task assignments, whereas the ODC, due to system heterogeneity and privacy or security constraints, accesses only high-level load information at the constellation level. When a user submits a task request, the ODC and NOCC jointly execute task planning and resource allocation. The ODC performs coarse-grained orchestration, determining which constellation should handle each task based on global load conditions, while the NOCC conducts fine-grained orchestration by assigning specific satellites and allocating precise bandwidth and computational resources. Upon receiving task instructions, the RSS and SCS establish inter-satellite links, transmit data, and perform on-orbit image processing. The processing results are then delivered to both the ODC and the user.}

\begin{table}[h!]
	\centering
	\caption{Parameter Settings}
	\begin{tabular}{c|c|c}
		\hline
		\textbf{Parameter} & \textbf{Value} & \textbf{Unit} \\
		\hline
		Number of Satellites  & 24 & / \\ 
		\hline
		Number of Ground Stations  & 1 $\sim$ 24 & / \\
		\hline
		Task Data Volume & 2 & GB \\
		\hline	
		Inter-Satellite Link Bandwidth  & 20 & Gbps\\
		\hline
		Satellite-to-Ground Link Bandwidth  & 1 $\sim$ 9 & Gbps\\
		\hline
		Processed Data Volume  & 20 & MB\\
		\hline
		Satellite Computing Capacity  & 300 & MB/s\\
		\hline
		Ground Server Cluster Computing Capacity  & 30 &GB/s\\
		\hline
	\end{tabular}
\end{table}

\begin{figure*}[!t]
	\centering
	\subfloat[Network Performance]{\includegraphics[scale=0.6]{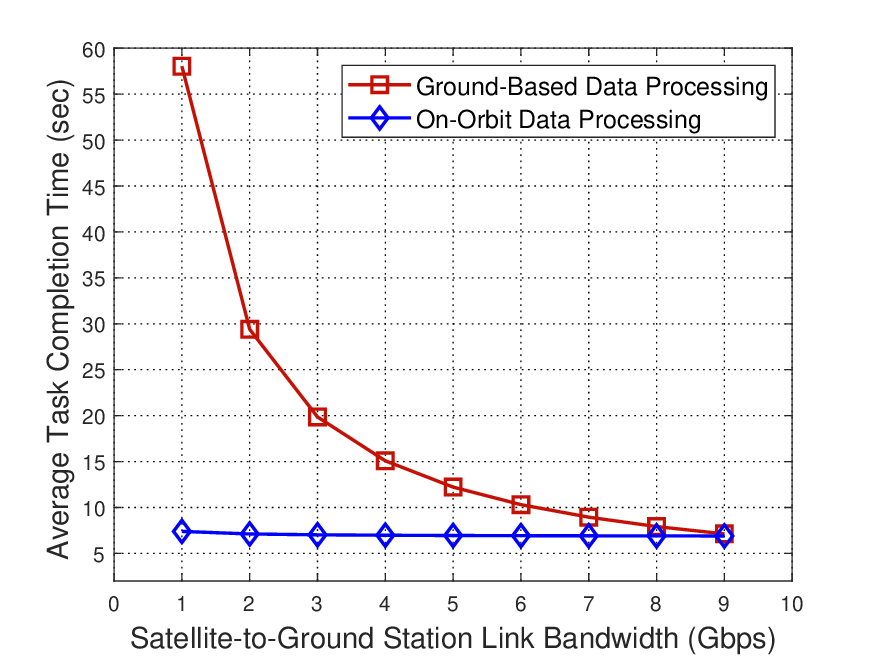}}
	\subfloat[Time Composition]{\includegraphics[scale=0.6]{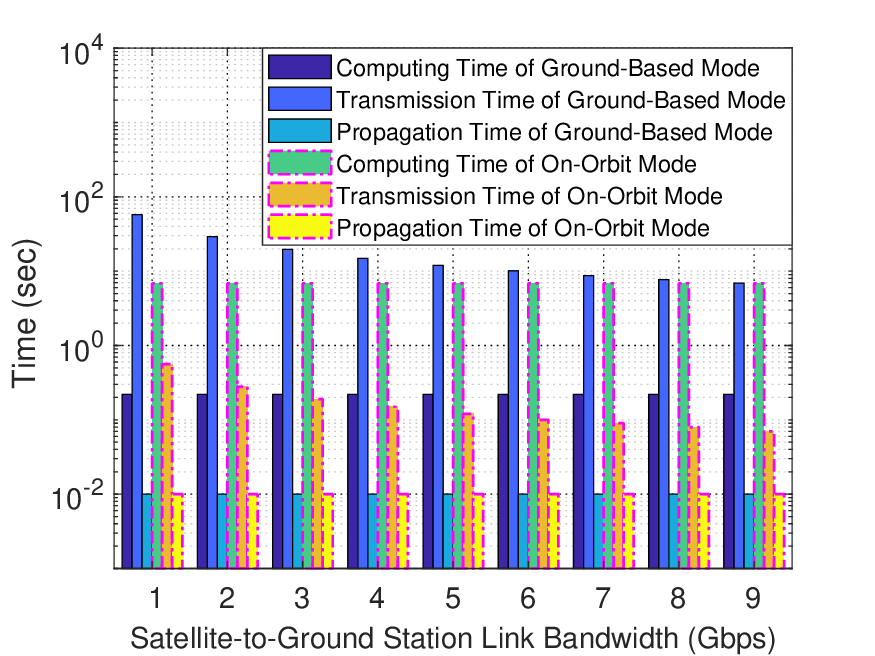}}
	\caption{System Analysis under Variable Satellite-Ground Link Rates.}
\end{figure*}

\begin{figure*}[!t]
	\centering
	\subfloat[Network Performance]{\includegraphics[scale=0.6]{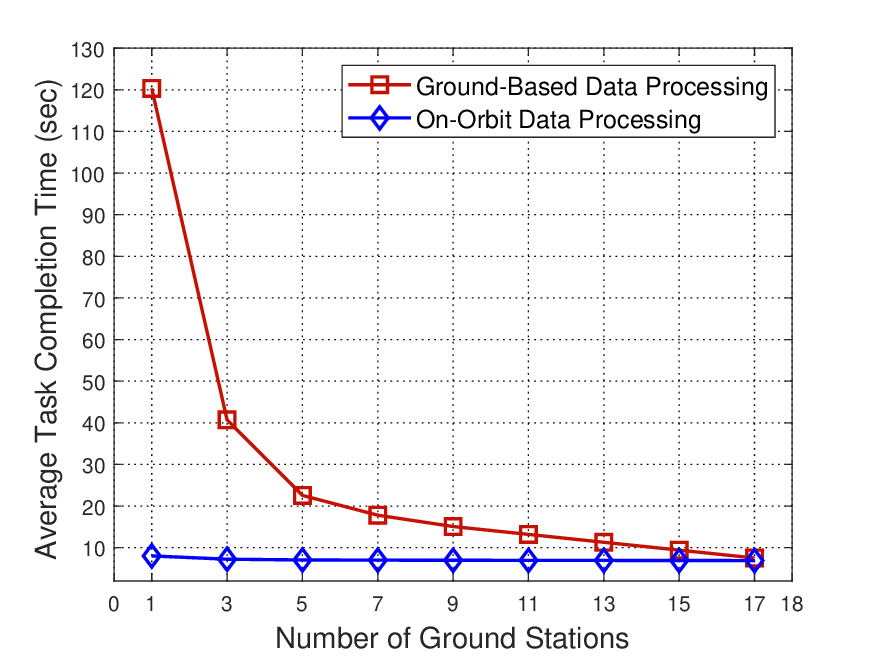}}
	\subfloat[Time Composition]{\includegraphics[scale=0.6]{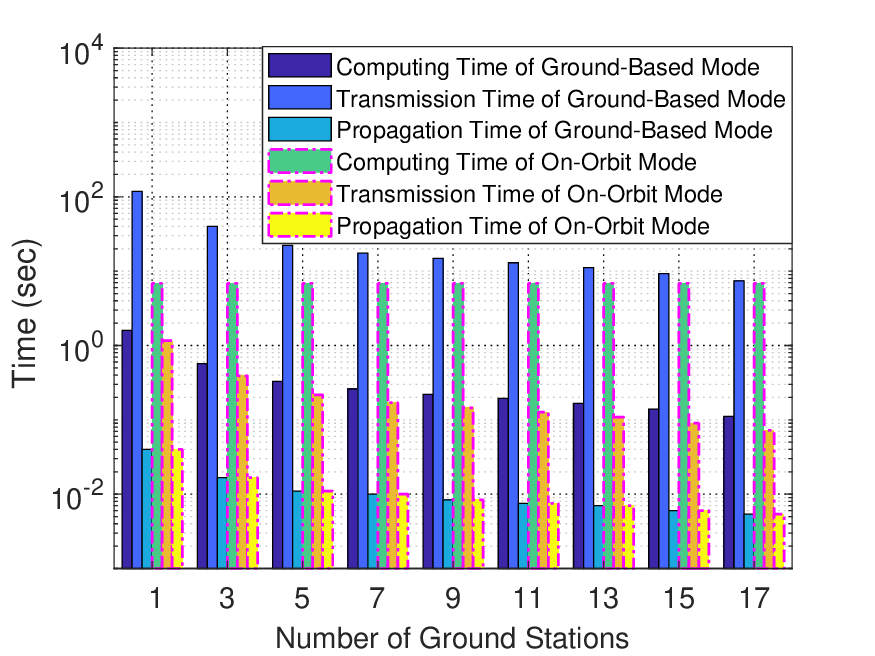}}
	\caption{System Analysis with Varying Ground Station Numbers.}
\end{figure*}

{To demonstrate the advantages of the proposed architecture over the traditional model, a simulation environment was constructed for quantitative analysis. Based on the Walker configuration used in Starlink, the simulation includes two orbital planes with a total of 24 satellites, as detailed in Table~1. Instead of using traditional metrics such as TOPS (Tera Operations Per Second) or TFLOPS (Tera Floating Point Operations Per Second) to quantify computational capability, we convert them into the equivalent amount of data processed per second. We focused our analysis on system performance as a function of satellite-to-ground link rate and the number of ground stations, as these two factors—outlined in the Introduction—are key constraints affecting task timeliness. The results are presented in Figure~4 and Figure~5, respectively.}

\begin{figure*}
	\centering
	\includegraphics[scale=0.5]{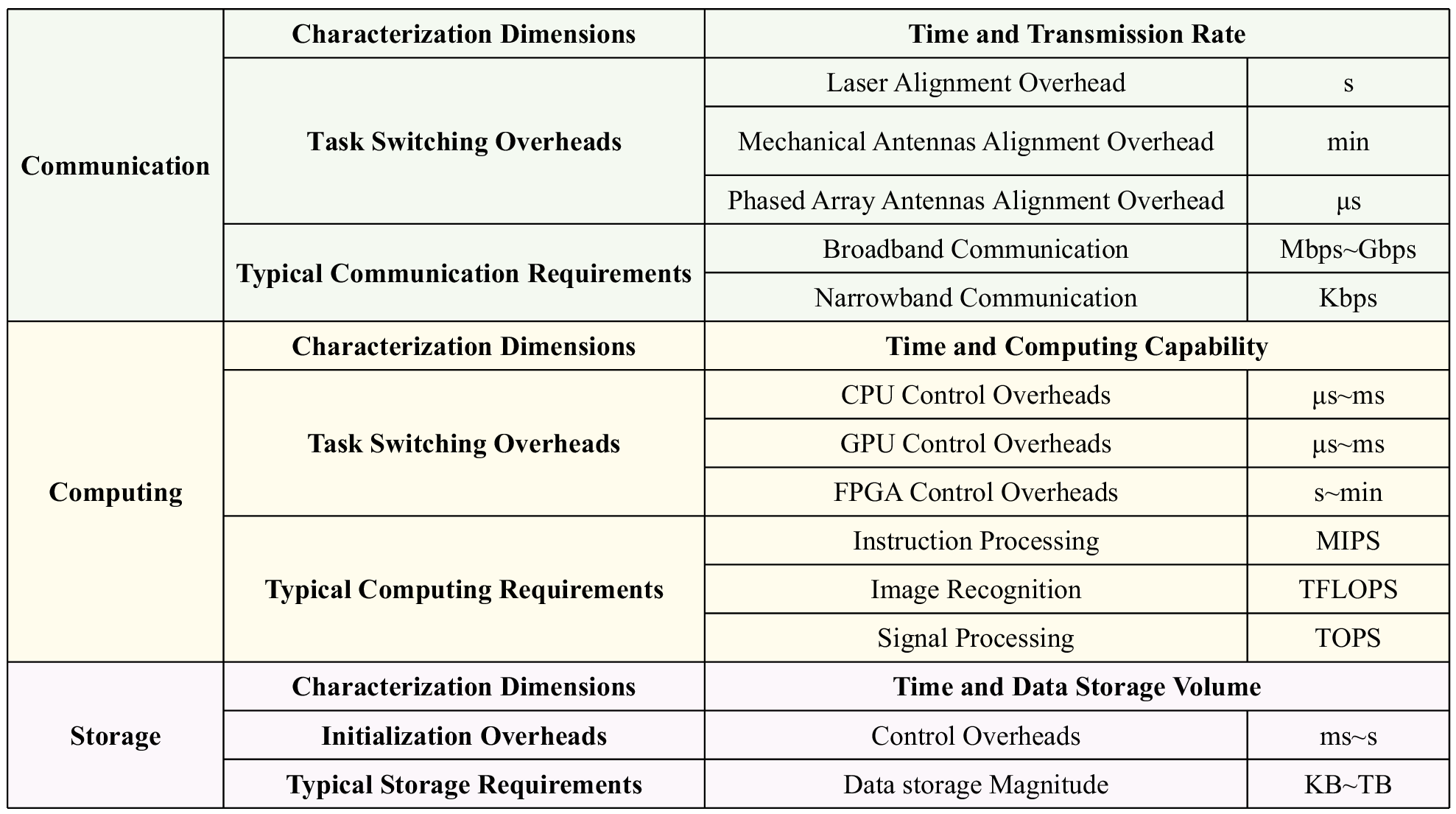}
	\caption{Characterization Dimensions for Heterogeneous Resources.}
\end{figure*}

{Figure~4 illustrates the impact of varying satellite-to-ground link rates on system performance, with the number of ground stations fixed at 9. As shown in Figure~4a, when the link rate is relatively low, the proposed space-based computing architecture achieves up to an 85\% reduction in task completion delay compared to traditional ground-based processing. This performance advantage is particularly relevant under typical real-world conditions (1 Gbps). Unless satellite-to-ground laser feeder link technologies can overcome weather-related limitations, achieving 10 Gbps remains challenging—only under such conditions can the traditional model achieve comparable performance. Figure~4b presents the composition of task completion time, including computation, transmission, and propagation components. Due to the large order-of-magnitude differences among these components across architectures, a logarithmic y-axis is used. Propagation time remains in the tens-of-milliseconds range and is significantly smaller than the other two components. Transmission time dominates in the ground-based model, whereas computation time is the primary contributor in the on-orbit model. As the link rate increases, transmission time decreases, allowing the performance of the ground-based model to gradually approach that of the on-orbit model.}

{Figure~5 illustrates the impact of varying the number of ground stations on system performance, with the satellite-to-ground link rate fixed at 4 Gbps. As shown in Figure~5a, increasing the number of ground stations reduces task completion time in the ground-based processing model, gradually approaching the performance of the on-orbit model. In practice, however, large-scale ground station deployment is challenging due to constraints related to policy, geography, and other factors, leading to significant regional disparities in task latency. Under such conditions, the on-orbit processing model offers distinct advantages. Figure~5b, similar to Figure~4b, presents the proportional breakdown of task completion time across computation, transmission, and propagation components. As the number of ground stations increases, transmission time decreases, thereby narrowing the performance gap between the ground-based and on-orbit processing models.}

\section{Challenges Analysis}

{Space-based computing network faces three key challenges: (1) the heterogeneous resource virtualization and state information synchronization; (2) the matching of multi-priority tasks with multidimensional resources; and (3) the fault detection and localization under extreme conditions. We begin by analyzing the fundamental differences between space-based and terrestrial networks, which give rise to unique challenges. We then elaborate on the underlying concepts in detail in the following subsections.}

\subsection{The Heterogeneous Resource Virtualization and State Information Synchronization}

\begin{figure*}[!t]
	\centering
	\includegraphics[scale=0.5]{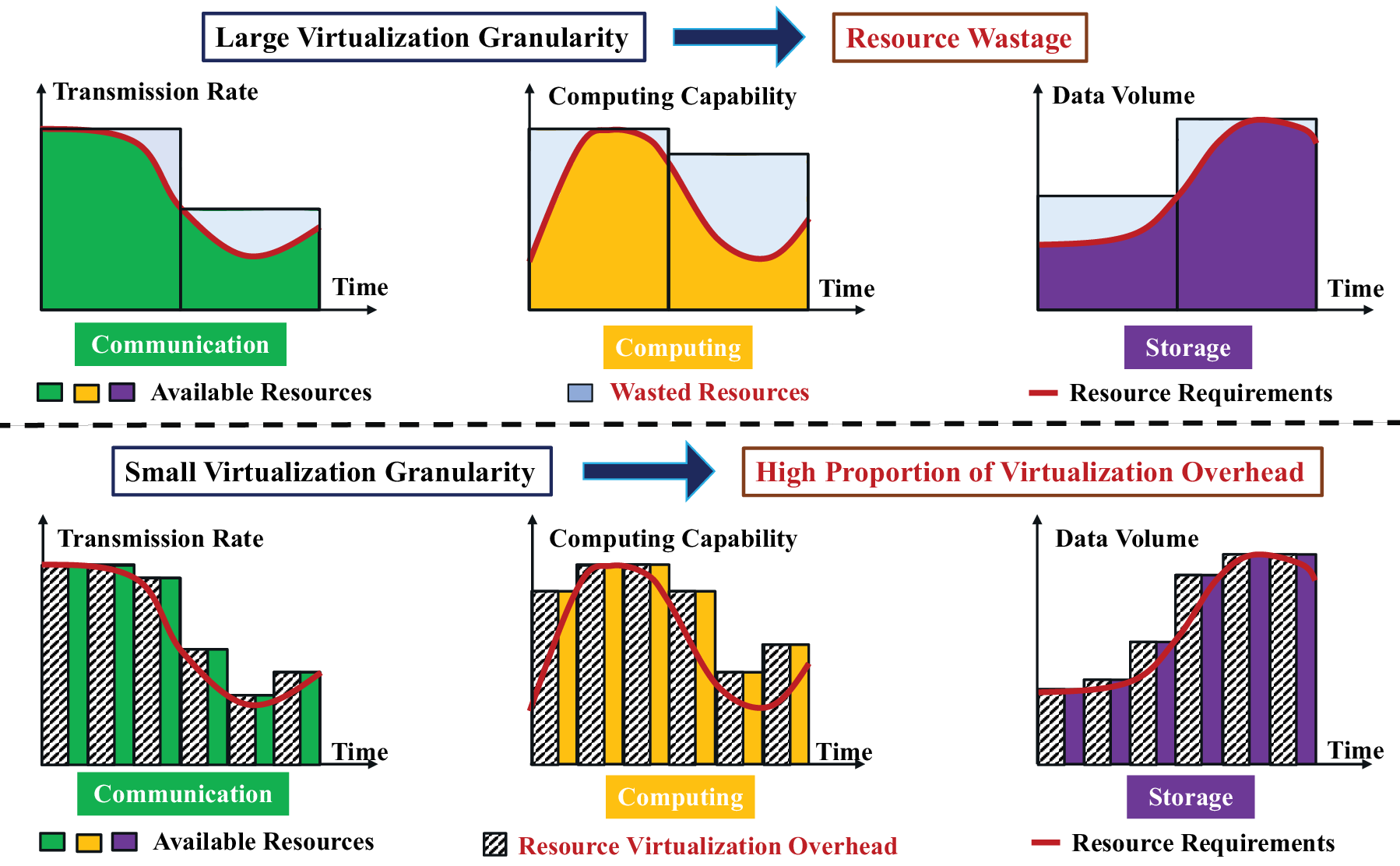}
	\caption{The Impact of Virtualization Granularity on Resource Utilization Efficiency.}
\end{figure*}

{Traditional ground-based computing networks typically rely on relatively homogeneous resources, with cloud centers using standardized computing and storage systems. In contrast, space-based computing networks must handle a wide variety of heterogeneous resources, including computational resources (e.g., CPUs, GPUs, and FPGAs), storage resources (e.g., RAM and ROM), and communication resources (e.g., different frequency-band links for space-to-ground and inter-satellite communications), which pose significant challenges for system-wide coordination. To address these challenges, virtualization technologies—capable of abstracting diverse hardware into standardized resource blocks—have emerged as a promising solution. Ground-based cloud centers, on the other hand, are fixed in location and equipped with ultra-high-speed fiber optic connections, enabling extremely fast information synchronization. In contrast, satellite networks in space experience high instability, constantly changing topologies, significant communication delays, and slow information synchronization. Therefore, heterogeneous resource virtualization and rapid state information synchronization are crucial for addressing the heterogeneity and dynamism of space-based computing networks.}

{The first step in resource virtualization is to analyze the characterization dimensions of various resources—that is, to evaluate the inherent overhead introduced by virtualization and to determine the typical magnitude of resource demand. On one hand, due to practical engineering constraints across different devices, understanding the task switching overhead (initialization cost) introduced during resource virtualization is critical. Otherwise, the overall system overhead may become prohibitively high. For example, when a single physical device is virtualized into five virtual resource blocks, time-division multiplexing (TDM) can lead to frequent task switching, resulting in substantial overhead. Here, we emphasize that our analysis primarily focuses on TDM-based virtualization technologies, and does not cover virtualization methods involving physical isolation at the hardware level. On the other hand, the typical resource demand determines the appropriate size of each virtual resource block. If the block size is poorly configured, resource utilization may remain low. As shown in Figure 6, we analyze the characterization dimensions of typical communication, computing, and storage resources, listing their task switching overheads and corresponding resource requirements.} 

{For communication resources, time and transmission rate are the primary characterization dimensions. In the case of laser communication, the time overhead for aligning the laser head to the target is typically on the order of seconds (s). In microwave and mmWave communication, there are two primary implementation methods: mechanical antennas and phased array antennas. Due to their larger size, which can affect satellite attitude, mechanical antennas are slower to rotate, with alignment times on the order of minutes (min). In contrast, phased array antennas support flexible beam-hopping techniques, resulting in alignment times on the order of microseconds (\textmu s). Communication systems can operate in two modes: broadband and narrowband. Broadband communication systems are primarily designed to support high-speed data transmission, with virtual resource block capacities typically on the order of megabits per second (Mbps) or gigabits per second (Gbps). In contrast, narrowband systems are mainly used for telemetry and control signaling, where the virtual resource block capacity generally falls within the kilobits per second (Kbps) range. }

{For computational resources, we take CPUs, GPUs, and FPGAs as representative examples. Virtualization technologies for CPUs and GPUs are relatively mature, with task-switching overheads ranging from microseconds to milliseconds, depending primarily on the hardware architecture and the management software. For example, in systems such as Kubernetes (K8s) \cite{k8s}, time-division multiplexing allows a single CPU to be virtualized into up to 1,000 instances that run alternately. In contrast, FPGAs incur significantly higher overhead due to the need for frequent compilation and circuit reconfiguration, with reconfiguration times potentially reaching the order of minutes. We also summarize the typical computational demands associated with these three types of devices. CPUs are well suited for instruction processing, and their virtual resource blocks generally need to support computing capabilities at the MIPS (Million Instructions Per Second) level. GPUs are mainly used for image processing, requiring virtual resource blocks to deliver TFLOPS (Tera Floating Point Operations Per Second) performance. FPGAs are commonly used for signal processing, with corresponding blocks needing to support TOPS (Tera Operations Per Second) level throughput. Given the wide variability in computational requirements across tasks, more precise specifications should be determined based on the application scenario. }

\begin{figure}
	\centering
	\includegraphics[scale=0.43 ]{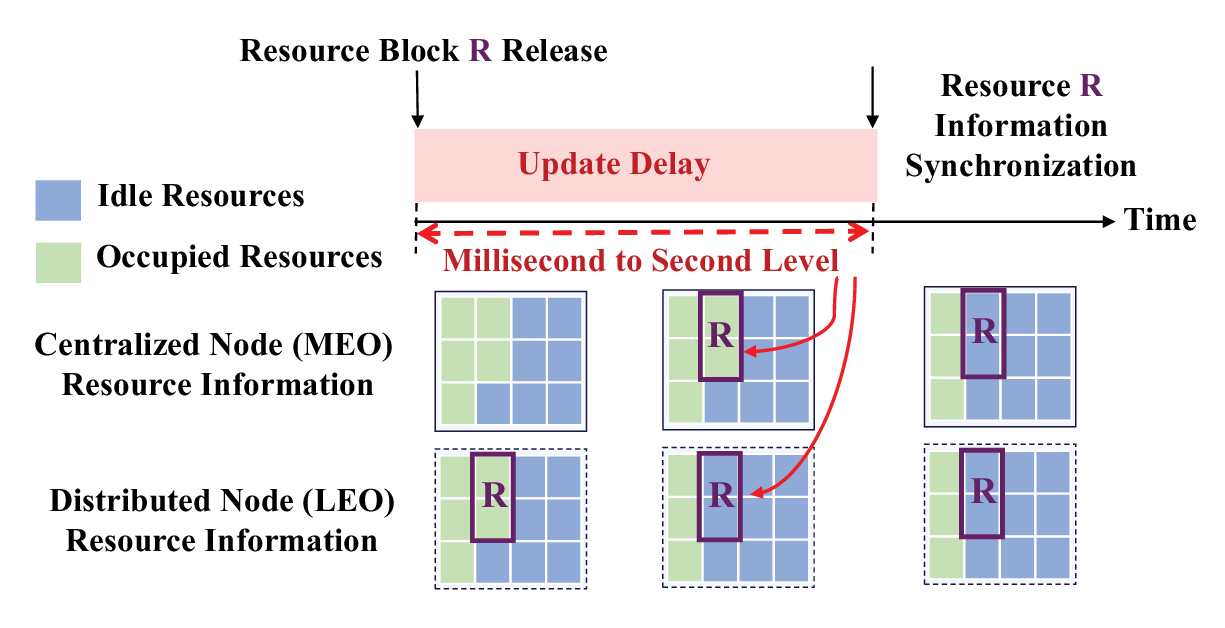}
	\caption{Update Delay of Resource Information.}
\end{figure}

{Unlike computing and communication resources, storage (SSD) does not incur task switching overhead caused by time-division multiplexing, as data remains intact unless explicitly deleted. Therefore, we only consider the initialization overhead, which is typically on the order of milliseconds. The amount of data that can be accommodated by virtualized storage resources spans a wide range, with the upper limit primarily determined by the capacity of the physical disk. It is worth noting that this analysis does not cover memory types such as RAM, as they are less relevant to satellite network management scenarios. In addition, due to the wide variety of resource types and virtualization technologies, an exhaustive enumeration is beyond the scope of this work. Instead, we offer a representative analytical perspective that may serve as a reference for researchers interested in applying it to more specific and well-defined scenarios.}

Figure~7 illustrates the impact of virtual resource block granularity on overall resource utilization. When the granularity is too coarse, resources tend to be underutilized, resulting in significant waste. Conversely, overly fine-grained virtualization can lead to excessive system overhead, thereby reducing effective utilization. Therefore, developing efficient multidimensional heterogeneous resource virtualization techniques and identifying an optimal balance in granularity is essential for maximizing resource efficiency.

{Another critical issue to consider is the synchronization of status information across all network entities. Space state information is difficult to update in real-time due to the varying orbits and continuous motion of satellites. The perception delay ranges from hundreds of milliseconds to several seconds, which can significantly impact the efficiency of resource allocation and release, as shown in Figure 8. Moreover, the volume of status information generated by a large number of satellites is substantial, creating a significant contradiction in the face of limited link bandwidth. Therefore, achieving real-time synchronization of status information is critical to ensuring the stable operation of the network.}

\subsection{The Matching of Multi-Priority Tasks with Multidimensional Resources}

{Traditional ground-based cloud data centers are equipped with large-scale server clusters and extensive fiber-optic infrastructure, providing abundant computational, storage, and communication resources. In contrast, most satellites weigh only a few hundred kilograms and can accommodate only a limited number of communication, computing, and storage payloads. The harsh space environment imposes stringent constraints on power supply—solar energy being the sole sustainable source—which fundamentally limits the upper bounds of computing and communication performance. Moreover, the bandwidth of inter-satellite communication links is typically limited to several tens of Gbps, which remains significantly lower than the data rates achievable with terrestrial fiber-optic networks. These constraints collectively underscore the severe resource limitations of space-based systems. Therefore, efficient task orchestration and resource scheduling are essential to meet the diverse Quality of Service (QoS) requirements of space applications.}

\begin{figure}
	\centering
	\includegraphics[scale=0.43 ]{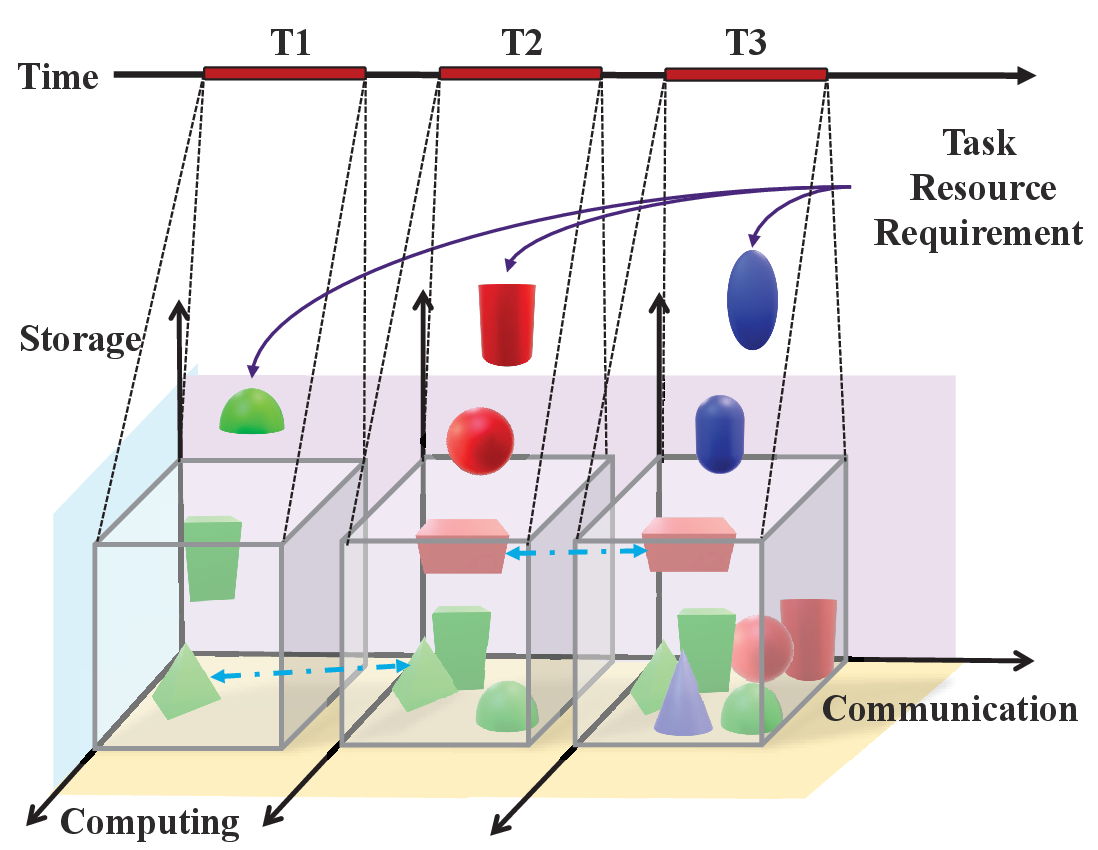}
	\caption{The NP-hard Multidimensional Temporal Knapsack Problem.}
\end{figure}

The differences in resource requirements for heterogeneous resources are substantial due to variations in task priority and processing stages, making orchestration extremely complex. We illustrate a typical multi-source remote sensing data fusion task, which consists of three stages: data access, image processing, and results distribution. The first stage of the task requires multiple continuous, stable high-speed data transmission links, with a rate in the Gbps range. In the second stage of the task, image processing in orbit requires the satellite to have robust computational and storage capabilities, with processing power on the order of hundreds of TOPS and storage capacity in the GB range. The final stage requires the satellite to rapidly distribute task processing results to multiple users across a wide area, with communication concurrency ranging from hundreds to thousands and a communication rate in the kbps range. In addition, the matching of multi-priority tasks with multidimensional resources is an NP-hard multidimensional temporal knapsack problem \cite{cacchiani2022knapsack} as shown in Figure 9, and no polynomial-time algorithm is currently known to solve it exactly. Therefore, exploring efficient orchestration of multidimensional heterogeneous resources for multi-priority tasks is crucial for ensuring the Quality of Service (QoS) of these tasks.

\subsection{The Fault Detection and Localization under Extreme Conditions}

{Traditional ground-based computing networks benefit from reliable power supplies, controlled environmental conditions (e.g., temperature and humidity), and redundant data protection mechanisms. As a result, they exhibit low failure rates and possess inherent self-healing capabilities. In contrast, the space environment is significantly more hostile: satellites may be constantly exposed to high-energy particle radiation and are vulnerable to various forms of cyber threats \cite{10836158}, both of which can induce system faults. Consequently, robust fault detection and localization under extreme conditions is essential for ensuring the reliability of space-based computing networks.}

\begin{figure}[h]
	\centering
	\includegraphics[scale=0.43]{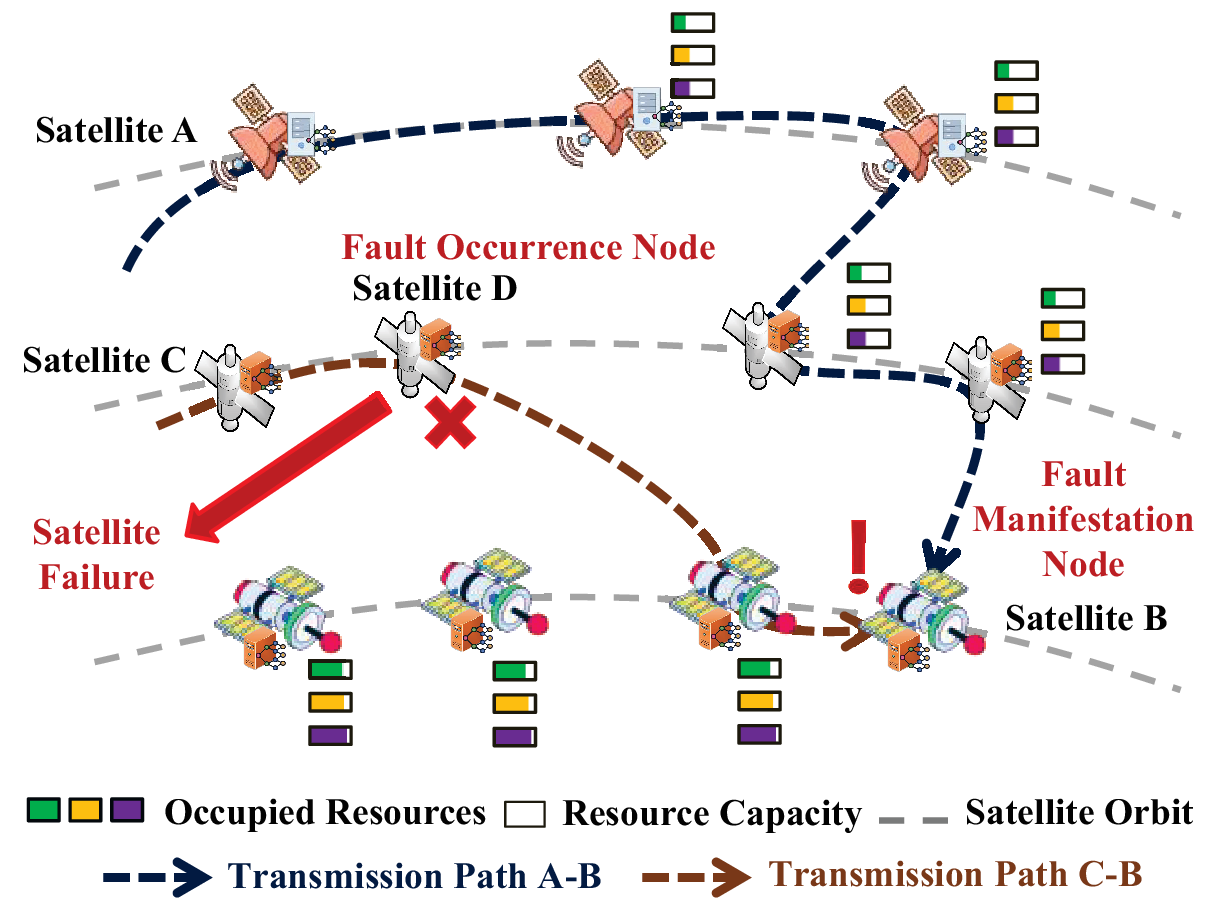}
	\caption{Satellite Network Failure.}
\end{figure}

{On one hand, modern satellites are becoming increasingly complex, equipped with a growing number of sensors but lacking sufficient historical fault data to support rapid and accurate diagnosis. On the other hand, as task execution typically involves numerous transmission and computation nodes, inconsistencies often arise between the point of fault origin and the point of detection, potentially misleading effective decision-making. As illustrated in Figure~10, Satellite~B is expected to process multisource data from both Satellite~A and Satellite~C. However, due to an unknown failure in Satellite~D, Satellite~B is unable to receive data from Satellite~C. In such cases, it becomes necessary to sequentially examine each node along the transmission path from Satellite~C to Satellite~B to locate the source of the failure. This issue is further exacerbated as task complexity increases and the network scale expands.}

\section{Key Technologies}

In this section, we discuss several key technologies that hold promise for addressing the challenges outlined in the previous section.

\subsection{The Heterogeneous Resource Virtualization and State Information Synchronization}

{Virtualization solutions designed for standardized resources in ground systems are clearly unsuitable for space scenarios involving large-scale heterogeneous resources. Therefore, tailored virtualization solutions must be developed for each resource type. The granularity of virtual resource blocks is a critical factor affecting overall system efficiency. It is necessary to explore methods for characterizing the relationship between granularity and system performance~\cite{laadan2010operating}, and to construct virtualization configuration profiles aimed at identifying the optimal granularity that balances resource utilization with system overhead. At the same time, reducing the system overhead introduced by virtualization has long been a key focus of research. Leveraging hardware collaboration to help reduce the overhead associated with virtualization is also a highly promising direction \cite{liu2024improving}. }

{The status information of each satellite must be deterministically transmitted to the NOCC to support task planning and resource allocation. To achieve this, discretizing the dynamic satellite network into a sequence of time-slot graphs is a necessary step \cite{9003306}. Subsequently, inter-satellite links can be fully leveraged to design a bandwidth allocation scheme for the entire network under a given latency threshold, ensuring timely information synchronization \cite{9761868}. To further reduce bandwidth consumption, transmitting only the delta (i.e., the change) of status data may be a promising strategy, which warrants further investigation.}


\subsection{The Matching of Multi-Priority Tasks with Multidimensional Resources}

{Task-resource matching is an NP-hard multidimensional knapsack problem whose complexity grows exponentially with the number of influencing factors. Tasks are generally categorized into pre-planned and emergency types. Pre-planned tasks are typically managed by orchestration algorithms at ground-based scheduling centers, where many well-established optimization theories can be applied \cite{10189398}. Although the computational complexity may be high, it remains acceptable due to flexible timing constraints. In contrast, for emergency tasks, timeliness becomes a critical performance metric, and the optimization objective often shifts toward balancing benefits and costs. Two typical strategies are employed in this context. The first involves searching for idle resources across the network. While this approach avoids disrupting existing task schedules, it may introduce excessive transmission delays if the resources are geographically distant, potentially violating time-sensitive requirements. The second strategy adopts proximity-based resource preemption. Although it enables rapid response, it may trigger cascading effects that compromise overall system efficiency. To address these challenges, researchers have proposed decomposing the global problem into simpler sub-problems using distributed techniques, or applying AI-based models trained offline and deployed onboard satellites to accelerate decision-making~\cite{10713242} \cite{ji2023cooperative}. Designing low-complexity, fast-response algorithms is essential for practical deployment.} From a task structure perspective, certain tasks consist of multiple interdependent subtasks organized in a graph structure \cite{10418555}\cite{10507245}. This necessitates efficient task partitioning to accelerate execution and minimize delays caused by subtask dependencies\cite{wu2019efficient}, especially in space networks. Additionally, the inherent uncertainty of task data volume can significantly impact orchestration performance~\cite{sun2024distributionally}\cite{11059567}. Typically, once a scheduling plan is formulated on the ground, satellite nodes execute their assigned sequences. However, due to geographic and environmental variations, task data volume may fluctuate. Building knowledge bases to support accurate forecasting represents a promising direction for achieving precise and adaptive resource management in future space-based systems.

\subsection{The Fault Detection and Localization under Extreme Conditions}

{Fault detection in satellite networks entails real-time awareness and intelligent analysis of abnormal states within satellite clusters. Existing methods can be broadly categorized into traditional empirical data processing and AI-driven knowledge-based approaches. The former relies heavily on historical cases, applying predefined parameters and decision thresholds for various fault types \cite{4512019}, but often suffers from limited diagnostic accuracy. In contrast, AI-based methods employ deep learning to extract feature representations of faults \cite{ibrahim2020machine} \cite{teng2023knowledge}, achieving higher detection accuracy but facing challenges in generalization. In the absence of large-scale datasets for training and evaluation, both approaches struggle to effectively handle rare or previously unseen failures. As a result, developing comprehensive, high-fidelity simulation environments and designing meta-learning frameworks with strong generalization capabilities are emerging as key research directions for future satellite fault detection systems \cite{Chang_2020}.}

{Fault localization in satellite networks is typically addressed through two categories of approaches: passive monitoring and active monitoring \cite{7471418}. Passive methods are inherently reactive, intervening only after a fault has occurred. These approaches utilize expert systems, artificial intelligence, and graph-theoretic models to infer fault locations by establishing mapping relationships among network nodes. Expert systems are generally based on decision tree theory and fuzzy logic theory, hierarchically traversing potential fault points using historical knowledge \cite{NAN200855}. AI-based approaches, similar to those employed in fault detection, leverage neural networks to extract features from past cases and predict likely fault locations \cite{10552813}. Graph-theoretic methods model fault propagation by constructing probabilistic relationships among nodes, functions, and their interdependencies to infer the origin of faults \cite{YU201744}. A primary limitation of passive monitoring is its inability to respond promptly, as it only activates post-failure. To overcome this, active monitoring has emerged, deploying multiple probing nodes across the network to proactively collect packet-level data and monitor node status \cite{QI2021108365}. By enabling end-to-end visibility, this approach provides a flexible and rapid diagnostic solution. However, active monitoring introduces considerable management complexity and incurs significant system overhead. As a result, abstracting the network structure and identifying critical nodes to balance diagnostic effectiveness and operational cost represent a promising direction for future research.}

\section{Conclusion}

In summary, this paper emphasizes the critical role of space information networks in the upcoming 6G era. The traditional model of transmitting space data to ground stations faces significant challenges, such as limited ground station deployment and constrained satellite-to-ground link rate. To address these issues, we propose a shift toward on-orbit data processing, marking the first steps toward a hierarchical space-based computing network. Our architecture integrates GEO, MEO, and LEO satellites with ground support systems, offering a comprehensive framework for their interaction. We identify three key challenges in space-based computing: heterogeneous resource virtualization and state information synchronization, matching of multi-priority tasks with multidimensional resources, and fault detection and localization under extreme conditions. These challenges demand innovative solutions to ensure high QoS and responsiveness. Additionally, we explore key technologies that are essential to overcoming these hurdles. Each of these areas presents unique complexities that require tailored solutions, distinct from those applied in terrestrial computing networks.

{The successful in-orbit operation of TSN-1 establishes a stable data fusion hub in medium Earth orbit (MEO) for remote sensing satellites equipped with heterogeneous sensor payloads. This facilitates improved on-orbit detection accuracy and enhances task timeliness. The mission underscores Tsinghua University's progress in space-based computing and lays the foundation for an open, shared experimental platform for future space network research.} Moving forward, we will continue to address the challenges in space-based computing networks, contributing Tsinghua’s expertise to the development of new productive forces. Our ultimate goal is to deeply integrate AI technology into the development of space networks, enabling autonomous on-orbit mission collaboration and heterogeneous multi-source information fusion, thereby building the intelligent brain of the space network.

\section*{ACKNOWLEDGEMENT}
\label{ACKNOWLEDGEMENT}

This work was supported by National Natural Science Foundation of China (62341130, 62341109, 62341106) and Shanghai Municipal Science and Technology Major Project.

\bibliographystyle{gbt7714-numerical}
\bibliography{myref}


%

\end{document}